\documentclass[prb,twocolumn,showpacs,amsmath,amssymb,aps]{revtex4}

\usepackage{graphicx}
\usepackage{dcolumn}
\usepackage{bm}

\begin{document}


\title{Effect of ferromagnetic contacts on spin accumulation in an all-metallic lateral spin-valve system:
Semiclassical spin drift-diffusion equations}
\author{Tae-Suk Kim$^{1,2}$, B. C. Lee$^{3}$, and Hyun-Woo Lee$^{1}$}
\affiliation{$^{1}$Department of Physics, Pohang University of Science and Technology,
    Pohang 790-784, Korea  \\
$^{2}$Asia Pacific Center for Theoretical Physics, Pohang University of Science and Technology,
    Pohang 790-784, Korea  \\
$^{3}$Department of Physics, Inha University, Incheon 402-751, Korea}

\date{\today}

\begin{abstract}
 We study the effect of the ferromagnetic (FM) contacts on the spin accumulation in the lateral
spin valve system for the collinear magnetization configurations.
When an additional FM electrode is introduced in the all-metallic lateral spin-valve system,
we find that the transresistance can be fractionally suppressed or very weakly influenced
depending on the position of the additional FM electrode, and relative magnitudes
of contact resistance and the bulk resistance defined over the spin diffusion length.
Nonlocal spin signals such as nonlocal voltage drop and leakage spin currents
are independent of the magnetization orientation of the additional FM electrode.
Even when the additional contact is nonmagnetic, nonlocal spin signals can be changed
by the spin current leaking into the nonmagnetic electrode.

\end{abstract}
\pacs{72.25.-b, 73.40.Gk}
\maketitle

\section{Introduction}
 Electrons are characterized by their quantized spin and charge.
In conventional electronic devices, only the charge degree of freedom has been employed 
for the control of the electron transport.
A new field of spintronics \cite{spinreview} was born from both experimental and theoretical efforts 
to harness the electron's spin degree of freedom in order to control the electric current in the devices. 
One of typical spintronic devices is the spin valve which
is a hybrid structure of ferromagnetic (FM) metal/nonmagnetic (NM) material/FM metal. 
The current passing through the spin valve
depends on the magnetization configuration of two FM metals. 
In the collinear case, usually more current flows through the spin valve in
the parallel configuration than in the antiparallel configuration.
Difference in resistance between the two is called magnetoresistance. 
In the noncollinear case or when two magnetization orientations are neither parallel nor antiparallel, 
the spin polarized current from one FM electrode exerts the spin torque
\cite{storque1,storque2,storque3} on the other FM electrode, and induces the magnetization dynamics. 
Examples of a spin valve are giant magnetoresistance (GMR) devices,\cite{gmr1,gmr2} magnetic
tunnel junctions,\cite{mtj} nanopillars,\cite{nanopillar} etc.

\begin{figure}[b!]
\includegraphics[width=7.0cm]{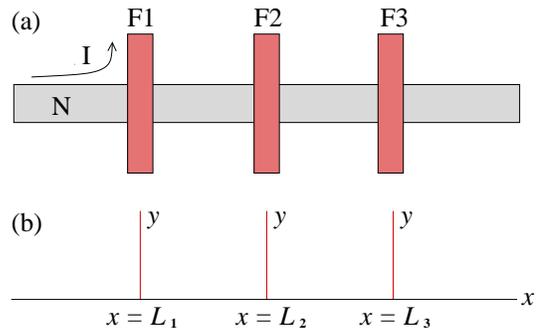}
\caption{(color online). Schematic display of the lateral spin valve
system with three ferromagnetic
 electrodes.
(a) Ferromagnetic electrodes are labeled as F1, F2, and F3 from left to right.
The base electrode is denoted as N.
Spins are injected from F1 to N by the spin-polarized current $I$ flowing from the left end of N into F1.
The electrode F1 is called the spin-injecting probe,
while F$i$ with $i\neq 1$ will be called the nonlocal (voltage) probe in this paper.
(b) The one-dimensional model geometry of the spin valve system in (a).
\label{3fmlead}}
\end{figure}

 In contrast with vertical spin valves, lateral spin valves are characterized by their multi terminal functionalities 
and so are more favorable for integration into semiconductor electronics. 
Due to increased spacing between terminals, efficient spin injection and detection have been a very hot issue. 
The spin injection and detection experiments in the two-terminal geometry are obscured by
other effects like anisotropic magnetoresistance, Hall effect, etc.
This defect was overcome by adopting the nonlocal spin valve geometry \cite{johnson0} similar 
to the schematic device structure in Fig.\ref{3fmlead}.
The original spin valve devices contain two FM electrodes (vs. three FM electrodes in Fig. \ref{3fmlead}) 
contacting the nonmagnetic base electrode.
In this lateral spin valve system, the spin transport was clearly observed with Al wires \cite{jedema} 
by spatially separating the spin current path from the charge current path and thereby removing other undesirable effects. 
The spin polarized current flows from the left of N (base electrode) into F1.
That is, spin polarized electrons are injected from F1 into base electrode N and is drained to the left of N. 
Due to asymmetry of two spin states in FM, the number of injected spin-up and spin-down electrons is different.
In addition to charge current in the left of N ($x<L_1$), diffusion
of injected spins generates spin current flowing to left and right
of N symmetrically. Pure spin current to the right of N was detected
\cite{johnson0, jedema,ji} with another FM electrode by measuring
the spin-dependent nonlocal voltage drop. The nonlocal spin
injection and detection technique was also used to observe
\cite{tinkham_SHE, otani_SHE} the (inverse) spin Hall effect in
diffusive nonmagnetic metallic strips. In these experiments, the
spatial separation of charge and spin currents as well as the
efficient spin injection are essential to observing the charge Hall
voltage induced by the spin current.

Recently experimental groups \cite{otani,otani2,johnson, wees}  studied the spin transport in the lateral spin valves 
with the three FM electrodes as shown in Fig.~\ref{3fmlead}. 
F1 is the spin-injection electrode (F$_{\rm si}$) as usual, while F2 and F3 are the nonlocal voltage probes 
located outside the charge current path.
While the spin polarized electrons are injected from F1 into N and are drained into the left of N, 
the transresistance is measured between F1 and F2 or between F1 and F3. 
In the former case, F2 is the spin-detecting electrode F$_{\rm sd}$ and F3 is the additional electrode  F$_{\rm a}$. 
In the latter case, the roles of F2 and F3 are switched. 
The important issue is how sensitively  the transresistance is affected by F$_{\rm a}$.

One experimental group \cite{otani} found that  F$_{\rm a}$ can drain the spin current and thereby 
significantly suppress the transresistance. 
They concluded that such additional FM electrode is relevant to the spin injection and accumulation in the multi terminal lateral spin valves.  
On the other hand, another group \cite{johnson} found that the transresistance is weakly affected by F$_{\rm a}$
even when the contact resistance between F$_{\rm a}$ and N is Ohmic. 
Moreover  the transresistance was observed \cite{johnson} to be independent of the magnetization orientation  of F$_{\rm a}$
 (parallel or antiparallel to that of F$_{\rm si}$).
They concluded that such additional FM electrodes are irrelevant to the spin injection and accumulation
in the multi terminal lateral spin valves.
 The existing experimental results seem to be contradictory to each other.

 Motivated by this experimental situation, we study theoretically the spin transport in the lateral spin valve 
with three ferromagnetic electrodes as schematically shown in Fig.~\ref{3fmlead}. 
We adopt the semiclassical spin drift-diffusion (SDD) equation \cite{johnson0, son} for the one-dimensional device structure
and study the mutual effect of FM electrodes on their nonlocal spin signals such as nonlocal voltage drop and leakage spin current. 
We find that the efficiency of the spin current leakage into FM electrodes depends on the relative magnitude of
junction resistance and the bulk resistance (defined over spin diffusion length) in FM and N electrodes. 
The voltage drop in $F_{\rm sd}$ is proportional to its leakage spin current with the proportionality constant 
given by the {\it effective} spin resistance which depends on the magnetization orientation of F$_{\rm sd}$.
Nonlocal spin signals are sensitive to the position of F$_{\rm a}$  
relative to the positions of F$_{\rm si}$ and F$_{\rm sd}$.
When F$_{\rm a}$ is located in between F$_{\rm si}$ and F$_{\rm sd}$, the transresistance can be either significantly or weakly 
affected by F$_{\rm a}$ depending on the relative magnitude of junction resistance and spin resistance.
The effect of F$_{\rm a}$ is weak when F$_{\rm a}$ is located outside the region between F$_{\rm si}$ and F$_{\rm sd}$.
Even though the magnitude of the spin current and nonlocal spin signals may be modified by F$_{\rm a}$, 
the flow direction of the spin current in the whole device is set by the magnetization orientation of F$_{\rm si}$
and so the nonlocal spin signals are independent of the magnetization orientation of F$_{\rm a}$.
This surprising result is already observed \cite{johnson} in experiments and is the direct consequence of no charge current 
in F$_{\rm a}$.  
Based on decoupling of charge and spin modes in the SDD equation and the Kirchhoff rules at the junction, 
we also show that the relationship between nonlocal spin signals and magnetization holds true even in realistic three-dimensional samples. 
These interesting properties in fact originate from zero charge current in nonlocal voltage electrodes.
Irrelevance of magnetization orientation of additional FM electrode to nonlocal spin signals implies that even
additional {\it nonmagnetic} electrode can modify nonlocal spin signals in the spin-detecting probe.
Our theoretical study  may be useful for clarifying the conflicting roles \cite{otani,otani2,johnson, wees} of an additional FM electrode 
in the lateral spin valve devices. 
In addition, our study is relevant to device applications, because the multi terminal functionality 
is essential for device applications of lateral spin valves.

 The rest of this paper is organized as follows.
In Sec. II, the spin drift-diffusion equation is briefly introduced
and the detailed algebras for the lateral spin valve with three FM
electrodes are included. The results of our work for spin valves
relevant to experiments are presented in Sec. III. In Sec. IV, our
work is summarized and its relevance to experiments is discussed.
Some algebraic details and interesting results are included in three
Appendices.

\section{Formalism}
 From now on we are going to confine our discussion to the collinear magnetizations of three FM electrodes
in spin valves and so we consider the spin polarized transport in a steady state. 
The noncollinear magnetizations go beyond the scope of our paper since the current flow in the noncollinear magnetizations 
generates the spin transfer torque and induces the magnetization dynamics.
In the collinear and diffusive transport, the spin drift-diffusion equations \cite{johnson0, son} have been very useful
for understanding phenomenologically the spin polarized transport in the spin valve systems. 
Later the SDD equations were derived \cite{valet_fert} from the semiclassical Boltzmann equation under the assumption 
that the spin diffusion length (SDL) is larger than the mean free path (MFP). 
Using the numerical solution of the spin dependent Boltzmann equation, the validity of the SDD equations was 
further extended \cite{penn_stiles} to the case when the SDL is comparable to the MFP. 
The SDD equations have been widely used for analyzing the spin injection experiments in
various device geometries.
The SDD formalism was also applied to the study of spin transfer torque \cite{brataas,kovalev,kovalev2} 
in the case of noncollinear magnetizations.

 The SDD equations in the collinear magnetizations
are written down for the spin-dependent electrochemical potential $\mu_{\alpha}$
and electric current density ${\bf j}_{\alpha}$.
Here $\alpha=\pm$ represents the spin-up ($+$) and spin-down ($-$) states, respectively.
The presence of the spin flip scattering in bulk mixes two spin states and the SDD equations
can be written down in a matrix form.
\begin{eqnarray}
\label{SDDE_3d}
\nabla^2 \begin{pmatrix} \mu_{+} \cr \mu_{-} \end{pmatrix}
  &=& \begin{pmatrix} \frac{1}{ D_{+} \tau_{+-} } & - \frac{1}{ D_{+} \tau_{+-} } \cr
               - \frac{1}{ D_{-} \tau_{-+} } & \frac{1}{ D_{-} \tau_{-+} }
      \end{pmatrix}
    \begin{pmatrix} \mu_{+} \cr \mu_{-} \end{pmatrix},  \\
{\bf j}_{\alpha}
  &=& \frac{\sigma_{\alpha}}{e} \nabla \mu_{\alpha}.
\end{eqnarray}
Here $D_{\alpha}$ is the diffusion constant for spin direction $\alpha=\pm$ and $\tau_{+-}$ is
the average spin-flip time for an electron from the spin direction $+$ to $-$.
$\sigma_{\alpha}$ is the conductivity for electrons with spin $\alpha$ and $e$ is the absolute value
of electron charge.

 The matrix differential equation for the electrochemical potential can be solved \cite{selman}
by analyzing the eigenvalues and eigenvectors of the matrix in the SDD equation.
One eigenvalue is $0$ and the corresponding eigenvector is
$\left( \begin{smallmatrix} 1 \\ 1 \end{smallmatrix} \right)$.
The other eigenvalue defines the spin diffusion length $\lambda$ and is given by the expression
\begin{eqnarray}
\label{SDL}
\frac{1}{\lambda^{2} }
  &=&  \frac{1}{ D_{+} \tau_{+-} } + \frac{1}{ D_{-} \tau_{-+} },
\end{eqnarray}
and its eigenvector is
$\left( \begin{smallmatrix} \sigma_{+}^{-1} \\ - \sigma_{-}^{-1} \end{smallmatrix} \right)$.
To find this form of the eigenvector, the Einstein relation as well as the detailed balance relation are
already invoked.
Obviously the first eigenvector (charge mode) does not discriminate between two spin states,
while the second one (spin mode) does.

 In this section we analyze the spin polarized transport in the spin valve system
based on the one-dimensional SDD equations. The device structure is
displayed in Fig.~\ref{3fmlead}(a), where the base electrode is
contacted to three ferromagnetic (FM) electrodes. Our primary goal
is to understand the mutual influence of the ferromagnetic
electrodes on the nonlocal spin signals such as the voltage drops
and the leakage spin currents. FM leads are labeled as F$i$ with
$i=1,2,3$ from left to right. The one-dimensional geometry,
corresponding to the device structure, is displayed in
Fig.~\ref{3fmlead}(b), where the junctions between the base
electrode and the FM leads are labeled as $x = L_i$ ($i=1,2,3$).

In experiments, the base electrode is nonmagnetic, but we are going to consider the case of magnetic
base electrode with its nonzero bulk spin polarization $\beta$.
Nonmagnetic case is recovered by a simple replacement $\beta = 0$.
The spin polarization (SP) in each FM lead is denoted as $\beta_i$ which is defined by the spin asymmetry
in the spin dependent conductivity $\sigma_{i\pm}$.
\begin{eqnarray}
\beta_i &=& \frac{\sigma_{i+} - \sigma_{i-}}{\sigma_{i+} + \sigma_{i-}}.
\end{eqnarray}
With the total conductivity $\sigma_i = \sigma_{i+} + \sigma_{i-}$, the spin-up and spin-down conductivities can be written as
\begin{eqnarray}
\sigma_{i\pm} &=& \frac{1}{2} (1 \pm \beta_i) \sigma_i.
\end{eqnarray}
For the base electrode, the spin polarization ($\beta$) in conductivity
and the spin-dependent conductivities ($\sigma_{\pm}$) are defined in a similar manner.

 When the spin polarized electrons are injected from F1 into the base electrode N
and is drained to left, the electrochemical potential in the FM leads
can be written as
\begin{eqnarray}
\label{ecp_FM}
\frac{1}{e} \begin{pmatrix} \mu_{i+} \cr \mu_{i-} \end{pmatrix}
  &=& \left[ \frac{I}{\sigma_1 A_1} y \delta_{i,1} - V_i \right] \begin{pmatrix} 1 \cr 1 \end{pmatrix}
    \nonumber\\
  && - I_i {\cal R}_i e^{-y/\lambda_i} \begin{pmatrix} (1+\beta_i)^{-1} \cr - (1-\beta_i)^{-1} \end{pmatrix}.
\end{eqnarray}
Here $\delta_{i,1}$ is the Kronecker delta function. The spin dependent current is determined by the equation
\begin{eqnarray}
I_{i\alpha} &=& A_i \frac{\sigma_{i\alpha}}{e} \frac{d}{dy} \mu_{i\alpha}.
\end{eqnarray}
The charge current is given by the expression
$I_{ci} = I_{i+} + I_{i-} = I \delta_{i,1}$ and flows only in F1 but not in F$i$ with $i\neq 1$.
The $i$-th FM lead is contacted to the base electrode at $x = L_i$. $A_i$, $\lambda_i$,
$\sigma_i$, and $\beta_i$ are the cross sectional area, the spin diffusion length, conductivity,
and bulk spin polarization in conductivity of the $i$-th FM lead, respectively.
${\cal R}_i$, defined by the relation
\begin{eqnarray}
{\cal R}_i &=& \frac{\lambda_i}{\sigma_i A_i},
\end{eqnarray}
is the resistance of the FM electrode over the spin diffusion
length. Due to an exponential decay of spin current, this definition
of resistance makes sense physically when discussing the spin current.
$V_i$ is the voltage drop at each ferromagnetic electrode far away
from the junction with the base electrode and is induced by the
nonequilibrium spin injection and diffusion. Note that the common
Fermi energy is dropped in writing the electrochemical potentials in
this paper, because the overall constant energy shift does not
change physics. The spin current in F$i$ $I_{i}^{s} = I_{i+} -
I_{i-}$ is given by the expression
\begin{eqnarray}
I_{i}^{s} &=& \beta_1 I \delta_{i,1} + I_i e^{-y/\lambda_i}.
\end{eqnarray}
The first term is the spin-polarized driving current, while the
second comes from the spin accumulation and diffusion. Though no
charge current flows in the region $x > L_1$, the spin current is
induced in the base electrode due to the spin injection,
accumulation and diffusion. The spin current decays exponentially
over the spin diffusion length and in turn leaks into the other FM
electrodes. $I_i$ measures the magnitude of this leakage spin
current at the interface between the base electrode and F$i$. The
leakage spin current also decays exponentially over the SDL in the
FM electrodes. The set of six unknown parameters $\{V_i, I_i\}$ are
to be determined by the Kirchhoff rules at the junctions. Note that
$V_i$ and $I_i$ are null for $i\neq 1$, when the injected current is
not spin polarized. $V_1$ can be nonzero for the tunneling barrier
even when the injected current is not spin polarized. Hence we may
call $\{V_i, I_i\}$ for $i\neq 1$ as the {\it nonlocal spin
signals}.

In the common base electrode, we have the electrochemical potential 
for spin-up($+$) and spin-down($-$) electrons
\begin{eqnarray}
\label{ecp_N}
\frac{1}{e} \begin{pmatrix} \mu_{+} \cr \mu_{-} \end{pmatrix}
  &=& \frac{I}{\sigma A} (x-L_1) ~ \theta(L_1-x) \begin{pmatrix} 1 \cr 1 \end{pmatrix}  \nonumber\\
  && - \sum_i J_i {\cal R} ~ e^{-|x-L_i|/\lambda}
     \begin{pmatrix} (1+\beta)^{-1} \cr - (1-\beta)^{-1} \end{pmatrix}.
\end{eqnarray}
Here $\theta(x)$ is the step function and the spin dependent current is computed from
\begin{eqnarray}
I_{\pm} &=& \frac{A\sigma_{\pm}}{e} \frac{d\mu_{\pm}}{dx}.
\end{eqnarray}
$A$, $\lambda$, $\sigma$, and $\beta$ are the cross sectional area, the spin diffusion length, conductivity,
and bulk spin polarization in conductivity of the base electrode, respectively.
${\cal R}$, defined by the relation
\begin{eqnarray}
{\cal R} &=& \frac{\lambda}{\sigma A},
\end{eqnarray}
is the resistance over the spin diffusion length in the base electrode.
The charge current $I_c = I_{+} + I_{-} = I \theta (L_1 -x)$ flows only at the section $x < L_1$
and the spin current $I_s = I_{+} - I_{-}$ is given by the equation
\begin{eqnarray}
I_N^{s} &=& \beta I \theta (L_1 -x)
+ \sum_i J_i \mbox{sgn}(x-L_i) ~  e^{-|x-L_i|/\lambda}.
\end{eqnarray}
The additional three unknown parameters $\{J_i\}$ are introduced for
the base electrode. $J_i$ represents the redistribution of spin
current in the base electrode due to its leakage into the voltage
probes (FM electrodes with $i=2,3$).

 The electrochemical potentials are constructed such that the charge current is conserved
at any junction in the device (charge continuity equation).
No net charge current flows to the right of $x=L_1$ and the currents for spin-up and spin-down
electrons are equal in their magnitude and opposite in their sign or flowing direction.
This symmetry in spin current is strictly obeyed in the nonlocal spin valve geometry even
in the presence of the multiple ferromagnetic electrodes to the right of the spin-injecting
FM electrode F1. Obviously the spin currents $I_i$ and $J_i$ are modified by the presence of
other FM electrodes, which is our main research interest in this work.

There are nine unknown parameters $\{V_i, I_i, J_i\}$ ($i=1,2,3$),
which should be determined by the boundary conditions or Kirchhoff
rules at the junctions. As already mentioned in the above, the
electrochemical potentials are constructed such that the charge
current is conserved. In our work, the spin flip scattering is
neglected at the interface, but is taken into account in bulks. In
this case, the spin current is conserved at each junction and the
following relations are obtained
\begin{eqnarray}
\label{rule_I}
J_i &=& \frac{1}{2} (\beta - \beta_1) I ~\delta_{i,1} - \frac{1}{2} I_i.
\end{eqnarray}
We consider the case of dirty interface between the ferromagnetic
electrodes and the base electrode. Due to a finite value of
resistance at the interface, the electrochemical potential across
the junction is not continuous and should be determined by the Ohm's
law. \cite{valet_fert,selman,takahashi}
\begin{eqnarray}
\label{rule_V}
\frac{1}{e} \Delta \mu_{i\alpha} ~
  &=& I_{i\alpha} (y = 0^{+}) {\cal R}_{ti\alpha}.
\end{eqnarray}
Here $\Delta \mu_{i\alpha}$ is the difference of the electrochemical potentials at $x = L_i$ between
the base electrode and the F$i$ electrode.
${\cal R}_{ti\pm}$ is the spin dependent junction resistance between the base electrode and F$i$,
and is defined in terms of the spin polarization $\gamma_i$ of junction resistance.
\begin{eqnarray}
\label{R_ts}
{\cal R}_{ti\pm} &=& \frac{2{\cal R}_{ti}}{1 \pm \gamma_i}.
\end{eqnarray}
${\cal R}_{ti}$ is the total junction resistance or
${\cal R}_{ti} = {\cal R}_{ti+} {\cal R}_{ti-}/({\cal R}_{ti+} + {\cal R}_{ti-})$.
The clean or transparent contact can be recovered by a simple replacement ${\cal R}_{ti} = 0$.

For the sign of $\beta$'s (spin polarization), we are going to adopt the following convention.
When the spin-up(-down) electrons belong to the majority(minority) channel at the Fermi level,
the sign of $\beta$'s is positive.
On the other hand, the sign of $\beta$'s is negative when the spin-up(-down) electrons belong to the minority(majority) channel.
According to our convention, the sign of $\beta$'s is reversed under the magnetization reversal.
The same convention applies to the sign of $\gamma$'s which are introduced
to define the spin polarization in the resistance of the interface.

 After some algebra as detailed in Appendix \ref{dalgebra}, we find the expressions for $I_i, J_i$ and $V_i$,
which contain all the information about the spin polarized transport
in the one-dimensional spin valve.
\begin{eqnarray}
\label{spin_I}
\frac{I_i}{I}
  &=& - (\beta_1-\beta) \delta_{i,1} + G_{i1} [ (\beta_1 -\beta) R_1 + (\gamma_1 -\beta) R_{t1}],  \\
\frac{J_i}{I}
  &=& - \frac{1}{2} G_{i1} [ (\beta_1 -\beta) R_1 + (\gamma_1 - \beta) R_{t1} ],  \\
\label{drop_V}
\frac{V_i}{I}
  &=& - [ (\beta - \beta_1)^2 R_1 + (\beta^2 - 2\beta \gamma_1 +1) R_{t1}] \delta_{i,1} \nonumber\\
    && + [ (\beta_i - \beta) R_i + (\gamma_i-\beta) R_{ti} ] G_{i1}  \nonumber\\
    && \times  [(\beta_1 - \beta) R_1 + (\gamma_1 - \beta) R_{t1}].
\end{eqnarray}
$G_{ij}$ is the element of the matrix ${\bf G}$ defined in Eq.~(\ref{cond_matrix}) of Appendix \ref{dalgebra}
and has the dimension of conductance.
The other set of material parameters, $R_i$ and $R_{ti}$, are introduced in Appendix \ref{dalgebra} and
their definitions are repeated here for readers.
\begin{eqnarray}
R_i &\equiv& \frac{{\cal R}_i}{1 - \beta_i^2}, ~
R ~\equiv~ \frac{\cal R}{1 - \beta^2}, ~
  R_{ti} ~\equiv~ \frac{{\cal R}_{ti}}{1 - \gamma_i^2}.
\end{eqnarray}
These new material parameters need our special attention.
They have the dimension of resistance and deserve their own terminology.
They are already called  the {\it spin resistance} in the literature.
First of all, the spin resistance is introduced to simplify the algebra
as shown in Appendix \ref{dalgebra}.
As the above equations show, this spin resistance determines the nonlocal spin signals such as
the voltage drops and the leakage spin currents in the voltage probes.
More physical insights on the spin resistance are elaborated on in Appendix ~\ref{s_resist}.

Since we are interested in the nonlocal transport measurements, we focus on the leakage
spin currents and voltage drops in the voltage probes (F$i$ with $i\neq 1$).
The leakage spin current in the nonlocal voltage probes is given by the expression [Eq.~(\ref{spin_I})]
\begin{eqnarray}
\label{nonlocal_I}
I_i &=& G_{i1} [ (\beta_1-\beta) R_1 + (\gamma_1 - \beta ) R_{t1}] I.
\end{eqnarray}
This relation for the nonlocal spin current $I_i$ suggests that the
conductance matrix ${\bf G}$ contains all the information about the
mutual effect of nonlocal voltage probes. Since the conductance
$G_{i1}$ does not depend on the magnetization configuration of the
FM electrodes, the leaking spin current does not depend on the
magnetization orientation of voltage probes, but instead depends on
the magnetization orientation of the spin-injecting FM electrode
(F1) and the base electrode (if ferromagnetic, $\beta\neq 0$). This
important observation can be understood as follows. The flow
direction of spin current (the sign of $I_i$), in the base electrode
as well as in the FM electrodes, is obviously determined by the
magnetization configuration in the spin-injecting electrode. This
means the flow direction of spin current cannot be altered by the
change of magnetic configurations in nonlocal voltage probes. This
is due to the fact that the nonequilibrium spin current is generated
by the spin-injecting electrode, but not by nonlocal voltage probes.
Another important observation is that the magnitude of spin current
or $I_i$ is not modified under the magnetization reversal of
nonlocal voltage probes, which in fact derives from the symmetry in
the SDD equations. This property of $I_i$ derives from decoupling of
spin and charge modes in SDD equations as well as the zero charge
current in nonlocal voltage probes. Detailed analysis can be found
in Appendix ~\ref{3D-case}. The relation $J_i = - I_i/2$ simply
reflects the conservation of spin current at the interface between
the voltage probe and the base electrode.

 $V_1^{s}$ defined below is ubiquitous in the expressions of $I_i$ and $V_i$.
\begin{eqnarray}
V_1^{s} &=& [ (\beta_1-\beta) R_1 + (\gamma_1 - \beta ) R_{t1}] I.
\end{eqnarray}
We may call $V_1^{s}$ as the {\it spin potential} which is the source from the spin-injecting electrode
and drives the spin current in the spin valve.
The leakage spin current in nonlocal probes can be written as
\begin{eqnarray}
\label{nonlocal2_I}
I_i &=& G_{i1} V_1^{s}.
\end{eqnarray}
The spin current in the spin valve device can be expressed in terms of the spin potential $V_1^{s}$ and
the conductance matrix ${\bf G}$ as
\begin{eqnarray}
I_N^{s} &=& \beta I \theta(L_1-x) \nonumber\\
   && - \frac{V_1^{s}}{2} \sum_i G_{i1} \mbox{sgn}(x-L_i) e^{-|x-L_i|/\lambda}, \\
I_1^{s} &=& \beta_1 I (1 - e^{-y/\lambda_1}) + [G_{11} V_1^{s} + \beta I ] e^{-y/\lambda_1},  \\
I_i^{s} &=& G_{i1} V_1^{s} e^{-y/\lambda_i}, i \neq 1.
\end{eqnarray}

 The voltage drop in the nonlocal voltage probes is given by the expression [Eq.~(\ref{drop_V})]
\begin{eqnarray}
\label{noloca_V1}
V_i &=& [ (\beta_i - \beta) R_i + (\gamma_i - \beta) R_{ti} ] G_{i1}  \nonumber\\
    && \times [(\beta_1 - \beta) R_1 + (\gamma_1 - \beta) R_{t1}] I.
\end{eqnarray}
Note that the voltage drop can be written as the product of spin current and some sort of spin resistance as
\begin{eqnarray}
\label{nonlocal_V2}
V_i &=& [ (\beta_i - \beta) R_i + (\gamma_i - \beta) R_{ti} ] I_i.
\end{eqnarray}
$V_i$ is the effective measure of weighted averaging the spin-up and
spin-down electrochemical potentials in F$i$ and so depends on the
magnetization configuration of F$i$. In addition, $V_i$ can be
understood as a shift in the Fermi level in order to satisfy the
condition of zero charge current in the nonlocal voltage probes.
Look at Appendix \ref{s_resist} for details. Though the leakage spin
current $I_i$ in F$i$ is independent of the magnetization
orientations of all nonlocal voltage probes (parallel or
antiparallel to that of spin-injecting probe), the voltage drop
$V_i$ depends on the magnetization orientation of F$i$, the
spin-injecting probe and the base electrode, but not on that of
other voltage probes.

\section{Results}
\label{svalve_3}
 In this section our discussion is confined to the spin valve system with three ferromagnetic electrodes
and {\it nonmagnetic} ($\beta=0$) base electrode. The current $I$ is
injected from the left of nonmagnetic base electrode and is drained
to F1 (See Fig.~\ref{3fmlead}). Though the charge current is null to
the right of the contact between F1 and the base electrode, the
finite spin current is induced everywhere by the spin injection,
accumulation and diffusion. With two nonlocal FM electrodes (labeled
as F2 and F3) contacted with the nonmagnetic base electrode to the
right of F1, we want to study the mutual influence of two nonlocal
FM electrodes on their voltage drops and leakage spin currents or
non-local spin signals.

 With three FM electrodes, the dimension of conductance matrix ${\bf G}$ is $3\times 3$.
\begin{eqnarray}
{\bf G}^{-1}
  &=& \begin{pmatrix}
   r_1 &  \frac{1}{2} R f_3 & \frac{1}{2} R f_2 \cr
   \frac{1}{2} R f_3  & r_2 & \frac{1}{2} R f_1   \cr
   \frac{1}{2} R f_2  & \frac{1}{2} R f_1 & r_3
      \end{pmatrix}.
\end{eqnarray}
Here $r_i = R_i + R_{ti} + \frac{1}{2} R$ for $i=1,2,3$ and
$f_1 = e^{-L_{23}/\lambda}$, $f_2 = e^{-L_{13}/\lambda}$, and $f_3 = e^{-L_{12}/\lambda}$,
where $L_{ij} = |L_i - L_j|$ is the distance between the contacts of the $i$-th and $j$-th FM electrodes
with the base electrode.
After inserting the explicit expressions of $G_{i1}$ into Eq.~(\ref{nonlocal_I}) with $\beta = 0$ or
\begin{eqnarray}
\label{3lead_I}
I_i &=& G_{i1} V_1^{s}, ~~~V_1^s = ( \beta_1 R_1 + \gamma_1 R_{t1}) I,
\end{eqnarray}
we find the leakage spin currents $I_2$ and $I_3$ in the nonlocal electrodes
to be given by the expressions
\begin{eqnarray}
\label{3lead_I2}
I_2 &=& - \frac{IR}{2D} e^{-L_{12}/\lambda} ~ (\beta_1 R_1 + \gamma_1 R_{t1})  \nonumber\\
   && \times \left[ R_3 + R_{t3} + \frac{1}{2}( 1 - e^{-2L_{23}/\lambda} ) R \right],  \\
\label{3lead_I3}
I_3 &=& - \frac{IR}{2D} e^{-L_{13}/\lambda} ~
    (\beta_1 R_1 + \gamma_1 R_{t1}) ( R_2 + R_{t2} ).
\end{eqnarray}
Here $D$ is the determinant of the matrix ${\bf G}^{-1}$ and is given by the expression
\begin{eqnarray}
\label{D}
D &=& \prod_{i=1}^{3} \left( R_i + R_{ti} + \frac{1}{2} R \right)  \nonumber\\
  && - \frac{1}{4} R^2 \sum_{i=1}^{3} \left( R_i + R_{ti} + \frac{(-1)^{i+1}}{2} R \right) f_i^2.
\end{eqnarray}
Obviously the exponentially decaying factor can be extracted out as
$G_{i1} = e^{-L_{1i}/\lambda} g_{i1}$ ($i=2,3$), where $g_{i1}$ is negative.
As mentioned in the previous section, the leaking spin current ($I_2$ and $I_3$) does not depend
on the magnetization configuration of the nonlocal voltage probes (F2 and F3),
but only on the magnetization configuration of the spin-injecting electrode (F1).
This is clearly explained by the fact that the spin current is generated by the spin-injecting electrode.

 The nonlocal voltage drops $V_2$ and $V_3$ are related to their leakage spin currents as
\begin{eqnarray}
\label{3lead_V}
V_i &=& (\beta_i R_i + \gamma_i R_{ti} ) I_i, ~~~ i =2,3.
\end{eqnarray}
This relation suggests that the nonlocal voltage drop $V_i$ is proportional to the corresponding nonlocal
spin current $I_i$ and the proportionality constant is the effective spin resistance
which depends on the magnetization orientation of F$i$.
This resistance is the intrinsic material properties of the relevant FM electrode so that
the effect of the other FM electrode is completely embedded into the nonlocal spin current.
This means that we can discuss the effect of multiple FM electrodes on transresistance
in terms of the leakage spin current.

From Eqs.~(\ref{3lead_I2}), (\ref{3lead_I3}) and (\ref{3lead_V}), transresistance
($R_{s2} =V_2/I$ and $R_{s3}=V_3/I$) in F2 and F3 electrodes are reduced to the following forms.
\begin{eqnarray}
\label{3lead_tR2}
R_{s2} &=& -\frac{R}{2D} e^{-L_{12}/\lambda} ~
    ( \beta_1 R_1 + \gamma_1 R_{t1} )
    ( \beta_2 R_2 + \gamma_2 R_{t2} )   \nonumber\\
  &&  \times
     \left[ R_3 + R_{t3} + \frac{1}{2}( 1 - e^{-2L_{23}/\lambda} ) R \right],  \\
\label{3lead_tR3}
R_{s3} &=& -\frac{R}{2D} e^{-L_{13}/\lambda} ~
    ( \beta_1 R_1 + \gamma_1 R_{t1} ) ( R_2 + R_{t2} )  \nonumber\\
  && \times ( \beta_3 R_3 + \gamma_3 R_{t3} ).
\end{eqnarray}
We denote the transresistance in the absence of an additional FM electrode using the superscript as
$R_{s2}^{(0)}$ and $R_{s3}^{(0)}$.
The same notations with superscript will be used for nonlocal voltage drop and spin current.
 The effect of other nonlocal FM electrode on the transresistance can be quantified by computing
the ratio: $R_{si}/R_{si}^{(0)}$ with $i=2,3$.
It follows from Eq.~(\ref{3lead_I}) and (\ref{3lead_V}) that
\begin{eqnarray}
\label{ratio0}
\frac{R_{si}}{R_{si}^{(0)}} &=& \frac{V_i}{V_i^{(0)}} = \frac{I_i}{I_i^{(0)}}
 = \frac{G_{i1}}{G_{i1}^{(0)}}.
\end{eqnarray}
The effect of an additional FM electrode on the transresistance can be measured by how much the nonlocal
spin current is reduced or by the change of the conductance matrix ${\bf G}$
under the other FM electrode.

 Two important facts can be read off from Eqs.~(\ref{3lead_tR2}) and (\ref{3lead_tR3}).
(i) The transresistance of one FM electrode (say, F3) does not depend on magnetization orientation
of the other electrode (say, F2).
That is, the transresistance $R_{s3}$ in F3 does not depend on the spin polarization,
$\beta_2$ and $\gamma_2$, of F2.
In Ref. \onlinecite{johnson}, the transresistance was observed
to be independent of the magnetization orientation (parallel or antiparallel to F3) of the intervening
FM electrode F2, which is supported by our theoretical results.
However this fact does not necessarily mean \cite{johnson} that 
the observed transresistance is not influenced by the additional FM electrode.
The transresistance can be either significantly changed or weakly influenced by the presence of the additional
intervening FM electrode, depending on sample and material parameters as we shall show below.
(ii) The transresistance can be modified \cite{otani2} even when the additional contacting electrode is nonmagnetic.
The relative magnitude of the interface and bulk resistance (defined over the SDL)
plays an important role in determining the transresistance.
Irrespective of the magnetic or nonmagnetic nature of the intervening electrode, the transresistance
will be influenced only by spin resistance and the interface quality.

 According to Eq.~(\ref{ratio0}), the effect of an additional FM electrode on the transresistance
is equivalent to its effect on the nonlocal leakage spin current.
After the spin current is injected from F1 electrode, it will flow
into both directions in N and will leak into nonlocal probes. From
this perspective we can expect that the effect will be much stronger
when an additional FM electrode lies in between two (spin-injecting
and spin-detecting) FM electrodes than when it lies outside two
electrodes. In the former case, the nonlocal spin current in
spin-detecting probe will be reduced proportionally by the amount of
spin current drained into an intervening electrode. In the latter
case, the injected spin current leaks into the spin-detecting probe
first and then into an additional FM electrode, so that the effect
will be weaker. Mathematically this difference between two cases
comes from the asymmetry between Eq.~(\ref{3lead_I2}) and
Eq.~(\ref{3lead_I3}). Under the index exchange $2\leftrightarrow 3$,
$I_2$ and $I_3$ are inequivalent due to the additional term
$(1-e^{-2L_{23}/\lambda})R/2$ in $I_2$.

To be more quantitative, let us consider $I_2$ when $f_i^2$'s in Eq.~(\ref{D})
are all much less than a unity.
This is a good approximation in all-metallic lateral spin valve systems since the spacing
between the electrodes is comparable to the SDL which is of the order of few hundred nanometers.
Under this approximation, we can readily show that
\begin{eqnarray}
I_2 &\simeq& I_2^{(0)}.
\end{eqnarray}
Therefore, $R_{s2} = V_2/I$ is very weakly influenced by the FM electrode F3 and
the transresistance is almost the same as that in the absence of the electrode F3.
That is, the transresistance is not much changed by the additional electrode (F3)
when it is contacted to the outside of F1 (spin current injected) and F2 (spin current detected).
However, the effect of an additional FM electrode F3 cannot be neglected if F2 and F3
are closer to each other than the SDL. So much for this case.

 We now focus on the case when an additional FM electrode lies in between the spin-injecting and
spin-detecting electrodes. That is, we study the effect of F2 on the
nonlocal spin signals for F3. In the absence of the intervening FM
electrode F2, $R_{s3}^{(0)}$ is \cite{takahashi,bclee}
\begin{eqnarray}
R_{s3}^{(0)} &=& - \frac{R}{2D_0} e^{-L_{13}/\lambda} ~
    ( \beta_1 R_1 + \gamma_1 R_{t1} ) ( \beta_3 R_3 + \gamma_3 R_{t3} ),  \\
D_0 &=& \left( R_1 + R_{t1} + \frac{1}{2} R \right) \left( R_3 + R_{t3} + \frac{1}{2} R \right) \nonumber\\
    && - \frac{1}{4} R^2 e^{-2L_{13}/\lambda}.
\end{eqnarray}
Note also that $R_{s3}^{(0)}$ can be obtained from $R_{s3}$ by taking the limit $R_{t2} \to \infty$
or when the second intervening F2 is effectively decoupled from the nonmagnetic base electrode.
The effect of the second intervening electrode F2 on the transresistance $R_{s3}$ can be quantified
by computing the ratio $R_{s3}/R_{s3}^{(0)}$,
\begin{eqnarray}
\label{ratio}
\frac{R_{s3}}{R_{s3}^{(0)}}
  &=& \frac{D_0}{D} (R_2 + R_{t2}).
\end{eqnarray}
Below this general relation will be reduced to the simple forms case by case.

 In order to provide some physical insights, let us consider the case when $f_i^2 \ll 1$.
We find the simple form of $R_{s3}/R_{s3}^{(0)}$
\begin{eqnarray}
\label{ratio2}
\frac{R_{s3}}{R_{s3}^{(0)}}
  &\simeq& \frac{R_2 + R_{t2}}{R_2 + R_{t2}+\frac{1}{2}R}.
\end{eqnarray}
The reduction of $R_{s3}$ stems from the leakage of spin current into the intervening electrode F2.
The efficiency of spin leakage into F2 is quantified by the relative magnitude of the serial resistance
$R_2 + R_{t2}$ in F2 and the resistance $R$ in base electrode over the spin diffusion length.
We can understand qualitatively the results of Eq.~(\ref{ratio2}) as
follows. Spins are injected from F1 into the base electrode N and in
turn diffuse into left and right of N. That is, the spin current
flows in N and leaks into nonlocal probes. Just like charge
transport, the spin current at the junction with F2 will continue to
flow in N and also leak into F2. If the effective spin resistance
$R_2 + R_{t2}$ of F2 is much larger than the spin resistance $R$ of
N, the leakage into F2 will be negligible and the spin current will
mostly continue to flow in N. The leakage spin current $I_2$ is
larger (smaller) if the effective spin resistance $R_2 + R_{t2}$ of
F2 is smaller (larger), compared to the spin resistance $R$ of N.
Obviously the leakage into F2 reduces the spin current in N and in
turn reduces the leakage spin current $I_3$. The larger (smaller)
$R_2 + R_{t2}$ is, the larger (smaller) is $I_3$.

 The spin diffusion length (SDL) is of the order of a few hundred nanometers (nm) in nonmagnetic metals
and the SDL in FM metals is of the order of a few nm to a few tens of nm.
The resistivity depends on the sample quality such as the impurities, defects, etc.
Though SDL is two orders of magnitude different between FM and NM,
the relative magnitude of resistance ($R_F$: ferromagnetic metal, $R_N$: nonmagnetic metal)
defined over the SDL can be varied from device to device.
Roughly $R_N \geq R_F$ in the spin valve devices.
Usually the interface between the FM electrodes and the nonmagnetic base electrode is Ohmic ($R_t$),
but not in the tunneling regime.
In real materials, we have the following order in resistance: $R_N \geq R_F > R_t$.
For our theoretical study, we will consider both cases of Ohmic and tunneling interfaces as well as
other parameter regimes.

\subsection{Clean F/N interface}
 We consider the clean interface between the base electrode and the FM electrodes:
$R_i, R \gg R_{ti}$. To get the simple expression of $R_{s3}$, we take the limit $R_{ti} = 0$.
Suppose that the FM electrodes are the same material with roughly the same $R_i \simeq R_F$
for $i=1,2,3$. If the exponentially decaying factors ($f_i^2$) are negligible, we find the simple form
of the transresistance ratio
\begin{eqnarray}
\frac{R_{s3}}{R_{s3}^{(0)}}
  &\simeq& \frac{2R_F}{2R_F + R},  \\
R_{s3}^{(0)} &\approx& -\beta_1 \beta_3 \frac{2R R_F^2}{(2R_F + R)^2} ~ e^{-L_{13}/\lambda}.
\end{eqnarray}
When $R_F \ll R$, the transresistance will be strongly suppressed by the additional intervening FM
electrode.
On the other hand, the transresistance will be fractionally reduced when $R_F$ is comparable to $R$.
In the other extreme case of $R_F \gg R$, the transresistance won't be affected by the intervening
FM electrode.

 As noted in the above, the transresistance can be affected by the nonmagnetic electrode,
$\beta_2 = 0$ and $R_2 = R_N$.
Let us study this case in detail. In the clean limit of interface,
\begin{eqnarray}
\frac{R_{s3}}{R_{s3}^{(0)}}
  &\simeq& \frac{2R_N}{2R_F + R},  \\
R_{s3}^{(0)} &\approx& -\beta_1 \beta_3 \frac{2R R_F^2}{(2R_F + R)^2} ~ e^{-L_{13}/\lambda}.
\end{eqnarray}
Since $R_N = \rho_2\lambda_2/A_2$ with $\beta_2 = 0$, we obtain the similar result
as in the previous paragraph depending on the relative magnitude of $R, R_N, R_F$.

 If the contacts between the base electrode and F1, F3 are clean, but the contact with the intervening
electrode F2 is in the tunneling regime, the effect of an additional electrode on the transresistance
is negligible.

\subsection{Tunneling F/N interface}
  When the junction resistance is dominant compared to the resistance over the spin-diffusion length
in the FM lead and the base electrode, or when $R_{ti} \gg R_j, R$,
the expressions of the voltage drop, Eqs.~(\ref{3lead_tR2}) and (\ref{3lead_tR3}), are simplified as
\begin{eqnarray}
\frac{V_2}{I}
  &\approx& -\frac{R}{2} \gamma_1 \gamma_2 e^{-L_{12}/\lambda},  \\
\frac{V_3}{I}
  &\approx& -\frac{R}{2} \gamma_1 \gamma_3 e^{-L_{13}/\lambda}.
\end{eqnarray}
The voltage drop at each junction is not influenced by the presence of the other FM leads,
when the junctions lie in the tunneling regime.
In general, the expression of $V_3$ is not affected by the presence of the second FM lead
(additional FM lead) as far as the contact is in the tunneling regime.
When $R_{t2} \gg R_i$, $R_{s3} = R_{s3}^{(0)}$ so that the second FM lead is effectively disconnected from
the base electrode.

 When the accumulated spin is diffused efficiently into the second intervening FM lead,
its effect may not be negligible.
We still assume that the contacts with F1 and F3 lie in the tunneling regime.
Let us see the extreme case of a transparent contact of F2 electrode to the base electrode.
In this case, we may set $R_{t2}=0$ and the desired voltage drop is given by the expression
\begin{eqnarray}
\frac{R_{s3}}{R_{s3}^{(0)}}
  &=& \frac{2R_2}{2R_2 + R},  \\
R_{s3}^{(0)} &=& -\frac{R}{2} \gamma_1 \gamma_3 e^{-L_{13}/\lambda}.
\end{eqnarray}
That is, the transresistance can be changed by the second intervening electrode F2 if F2 is in clean
contact with the base electrode or if spin leakage into F2 is efficient.

\section{Discussion and Summary}
 Using the one-dimensional spin drift-diffusion equations,
we studied theoretically the mutual effect of ferromagnetic
electrodes on non-local spin signals (the leakage spin currents and
the voltage drops) in the lateral spin valve with three
ferromagnetic electrodes. We found the generic expression of the
leakage spin current, Eq.~(\ref{nonlocal_I}), and also a very simple
relation, Eq.~(\ref{nonlocal_V2}), between the nonlocal voltage drop
and the leakage spin current.

Eq.~(\ref{ratio0}) tells us that the effect of an additional electrode on the transresistance can be
discussed in terms of the leakage spin current and in turn in terms of the conductance matrix.
The measured non-local spin signals depend on the position of an additional FM electrode
relative to the spin-injecting and spin-detecting electrodes.
When the additional electrode lies outside of the two FM electrodes, non-local spin signals are found to be weakly influenced 
due to the exponentially decaying spin coherence over the SDL.
On the other hand, when it is located in between the two FM electrodes,
the non-local spin signals can be strongly modified provided the junction resistance is lower than or comparable to  
the spin resistance defined over the spin diffusion length in the FM electrodes and the nonmagnetic base electrode. 
If the junction resistance is high, the non-local spin signals are weakly modified 
even when the additional FM electrode is located in between the two FM electrodes
The most general expression for the transresistance ratio is given by Eq.~(\ref{ratio}).
In general, the non-local spin signal is not much modified
when the additional electrode is in tunneling
contact with the base electrode, but is fractionally reduced when the contact is Ohmic.
We also found that the non-local spin signals are independent of the magnetization orientation of
the additional FM electrode, which agrees with the experimental observation.\cite{johnson}
This result suggests that even the intervening {\it nonmagnetic} electrode can change
non-local spin signals, which was already observed \cite{otani2} experimentally.

 Since our theoretical study is based on the one-dimensional device structure,
some care is needed when we try to apply our theoretical results to interpretation of experimental data.
Strictly speaking, the experimental spin valve structure is not one-dimensional in terms of the current distribution.
Harmle {\it et al.} numerically showed \cite{hamrle} that the nonlocal voltage drop
depends strongly on the spatial distribution of the spin-polarized current.
The one-dimensional approximation is valid when the current is uniformly distributed through the contact.
When the contact is clean between the FM electrode and the base nonmagnetic electrode,  
the current flow may well not be uniform through the interface \cite{johnson} and may be short circuited.
In this case, the nonlocal spin signals may deviate from its theoretical estimate based
on one-dimensional SDD equations.
Keeping these restrictions in mind,
let us apply our theoretical results to two experimental works. \cite{otani, johnson}

For numerical estimation ($R_s/R_{s0}=R_{s3}/R_{s3}^{(0)}$ in this section),
we take examples of Co/Cu/Co and Py/Cu/Py
lateral spin valves and use the following sample size and material parameters.
The thickness and width of the nonmagnetic base electrode
are taken as 80 nm and 300 nm, respectively. The width of all the
ferromagnetic layers is assumed to be the same as 100 nm. The
separation between nearest ferromagnetic layers is taken as 200 nm,
which gives 300 nm of center-to-center distance. We use material
parameters measured at low temperatures. The parameters for Cu are
$1/\sigma_{\rm Cu}=6 \times 10^{-9}\; \Omega$m\cite{Bass} and
$\lambda_{\rm Cu}=1 \; \mu$m.\cite{jedema} For Co, we use
$\beta_{\rm Co}=0.46$,\cite{Bass} $\gamma_{\rm
Co/Cu}=0.77$,\cite{Bass} $1/\sigma_{\rm Co}(1-\beta_{\rm Co}^2) =
7.5 \times 10^{-8} \; \Omega$m,\cite{Bass} $\lambda_{\rm Co}=59$
nm,\cite{Fert} and $R_{\rm Co/Cu}A/(1-\gamma_{\rm Co/Cu})^2=0.52
\times 10^{-15}\; \Omega$m$^2$.\cite{Bass} For Py, we take
$\beta_{\rm Py}=0.73$,\cite{Bass} $\gamma_{\rm
Py/Cu}=0.70$,\cite{Bass} $1/\sigma_{\rm Py}(1-\beta_{\rm Py}^2) =
15.9 \times 10^{-8} \; \Omega$m,\cite{Bass} $\lambda_{\rm Py}=5.5$
nm,\cite{Bass} and $R_{\rm Py/Cu}A/(1-\gamma_{\rm Py/Cu})^2=0.54
\times 10^{-15}\; \Omega$m$^2$.\cite{Bass}

For the Co/Cu/Co spin valve, $R/2=125$ m$\Omega$, $R_2=150$  m$\Omega$, and $R_{t2}=17$ m$\Omega$ are obtained.
The estimated spin signal is reduced to the value $R_s/R_{s0}=0.68$ by the intervening F2 electrode.
$V_2/I$ is also reduced by the factor 0.87 due to F3 electrode.
For the Py/Cu/Py spin valve, we have $R/2=125$ m$\Omega$, $R_2=29$ m$\Omega$, and $R_{t2}=18$
m$\Omega$. The reduced spin signal $V_3/I$ by the F2 electrode
is $R_s/R_{s0}=0.46$. $V_2/I$ is reduced by the factor 0.87 due to F3 electrode.

Since the SDL of Co is rather long, $R_2$ is comparable to $R/2$ in the Co/Cu/Co case
and $R_s/R_{s0}$ is large.
Since, in the Py/Cu, $R/2$ is bigger than $R_2$ and $R_{t2}$, $R_s/R_{s0}$ is small.
The rather significant reduction of estimated $V_2/I$
in both cases stems from our choice of the long SDL of Cu at low temperatures.
The long SDL means that the chemical potential splitting between
opposite spin directions, though exponentially decaying, remains significant
up to the position of the F3 electrode.
The significant leakage of spin currents into F3 results in reduction of the spin signal.
At room temperature, the SDL of the base electrode (nonmagnetic metal) is a few hundreds nanometers
such that the reduction of $V_2/I$ by the F3 electrode is only a few percents.
For the experimental conditions in Refs.
\onlinecite{otani} and \onlinecite{johnson}, $R/2$ is comparable to
$R_2 + R_{t2}$ and we can estimate theoretical value of $R_s/R_{s0}$, Eq.~(\ref{ratio2}):
$R_s/R_{s0} \simeq 0.5$ although the observed $R_s/R_{s0}$ is smaller for Ref.
\onlinecite{otani} and is close to a unity for Ref. \onlinecite{johnson}.

 As pointed out in Ref.~\onlinecite{johnson}, the contact between the Permalloy electrode and the base Ag wire
is very clean and the point injection and detection of current is suggested. 
In this case, the current distribution in the devices  may well be nonuniform so that 
our one-dimensional theory cannot be straightforwardly applied.
We believe that the nonuniform current distribution is the main reason why some of our theoretical estimates are 
in poor agreement with the results of Ref.~\onlinecite{johnson}.
We may discuss the relevance of the device dimensionality based on the effective spin resistance.
According to our theoretical analysis, nonlocal spin signals are determined by the relative
magnitude of junction resistance and spin resistance in FM and NM electrodes.
This relevant resistance is defined under the assumption that the current distribution
is uniform in the device.
When the current distribution is not uniform as in real devices,
we may still be able to define the spin resistance using the effective cross sectional area
which is smaller than the geometrical cross section of the sample.
Nonuniform current distribution tends to increase
junction resistance as well as spin resistance,
and will modify the magnitude of nonlocal spin signals.
This may be one of the reasons for the discrepancy between two experimental results.\cite{otani,johnson}

Nonlocal spin signals (the leakage spin current and the voltage
drop) in one nonlocal FM electrode are shown not to depend on the
magnetization orientation (parallel or antiparallel) of the other
nonlocal FM electrode. We believe this symmetry of nonlocal spin
signals are robust against the sample dimensionality, though their
magnitude is sensitive to samples. The spin current is generated by
the spin-injecting FM electrode and so its flow direction cannot be
changed by the magnetization orientation of nonlocal FM electrodes.
In addition, the magnitude of spin current does not depend on the
magnetization orientation of nonlocal FM electrodes. This property
derives from both decoupling of spin and charge modes in the SDD
equations and zero charge current in nonlocal voltage probes. Hence
our conclusion about the relationship between nonlocal spin signals
and magnetization in nonlocal voltage probes won't depend on the
sample dimensionality and qualities. This point is demonstrated more
explicitly in Appendix~\ref{3D-case}.

Finally, we would like to discuss the properties of transresistance (TR) 
and (longitudinal) magnetoresistance (MR) in spin valves under magnetization reversal.
Obviously, both TR and MR are modified under magnetization reversal of two probing FM electrodes. 
Under magnetization reversal, TR changes its sign while MR changes its value.
Note that MR, in general, consists of the two contributions:
one part (background) remains the same but the other changes its sign under magnetization reversal. 
Let us consider the effect of an additional FM electrode (F$_{\rm a}$) on TR and MR. 
For the vertical spin valves, it won't be easy to implement F$_{\rm a}$.
So we consider TR and MR in the lateral spin valves with F$_{\rm a}$.
Usually MR is obscured by other effects in the lateral spin valves as mentioned before. 
However, with increased SDL, MR was successfully measured \cite{mr_cnt1,mr_cnt2} in the carbon nanotube and graphenes.
To measure MR in the lateral spin valves of Fig.~\ref{3fmlead}, F1 and F3 are both current and voltage probes.
Based on the results of Appendix~\ref{3D-case}, 
we can argue that MR should be independent of the magnetization orientation of F2, \cite{BNB_circuit}
because there is no charge current in F2 (an additional electrode).
Explicit calculation, \cite{tskim} using the SDD equations, confirms this claim. 
That is, both TR and MR are independent of the magnetization orientation of F$_{\rm a}$.
On the other hand, TR and MR depend on the magnetization orientation of ferromagnetic electrodes
through which the charge current flows.

\acknowledgments
This work was supported by the Korea Science and Engineering Foundation (KOSEF) grant funded by
the Korea government (MOST) (No. R01-2005-000-10303-0), by the SRC/ERC program of MOST/KOSEF (R11-2000-071)
and by POSTECH Core Research Program.

\appendix

\section{Details in algebra}
\label{dalgebra}
 In this Appendix we show that the algebraic manipulation can be highly simplified by a proper
definition of material parameters and by the vector and matrix notations.
 Kirchhoff rules lead to the constraints given by Eqs.~(\ref{rule_I}) and (\ref{rule_V}).
Eq.~(\ref{rule_V}) can be written down explicitly leading to the
following six relations ($i=1,2,3$)
\begin{eqnarray}
\label{spin_R}
&& \pm \sum_{j} \frac{J_j {\cal R}}{1 \pm \beta} ~ e^{-|L_i - L_j|/\lambda}
  - \left[ V_i \pm \frac{I_i {\cal R}_i}{1 \pm \beta_i} \right] \nonumber\\
  && = \left[ ( 1 \pm \beta_1 ) I ~ \delta_{i,1} \pm I_i \right] \times
      \frac{{\cal R}_{ti}}{1 \pm \gamma_i}.
\end{eqnarray}
For algebraic convenience, we introduce new material parameters
\begin{eqnarray}
R_i &=& \frac{{\cal R}_i}{1 - \beta_i^2}, ~
R ~=~ \frac{\cal R}{1 - \beta^2}, ~
  R_{ti} ~=~ \frac{{\cal R}_{ti}}{1 - \gamma_i^2},
\end{eqnarray}
and $A_{ij} ~=~ e^{-|L_i - L_j|/\lambda}$.
These new material parameters highly simplify the complicated algebra as well as
determine the spin currents in the nonlocal voltage probes.
Then the voltage drop $V_i$ can be written as
\begin{eqnarray}
V_i &=& \mp I_i R_i ( 1 \mp \beta_i)
       \pm (1 \mp \beta) R \sum_{j} A_{ij} J_j  \nonumber\\
    && - (1 \mp \gamma_i) R_{ti} \left[ (1 \pm \beta_1) I \delta_{i,1} \pm I_i  \right].
\end{eqnarray}
Addition and difference of two $V_i$'s lead to
\begin{eqnarray}
V_i &=& -(1 - \gamma_1 \beta_1) R_{t1} I ~ \delta_{i,1}
      + ( \beta_i R_i + \gamma_i R_{ti} ) I_i  \nonumber\\
    && - \beta R \sum_{j} A_{ij} J_j,
\end{eqnarray}
and
\begin{eqnarray}
(R_i + R_{ti} ) I_i - R \sum_{j} A_{ij} J_j
  &=& (\gamma_1 - \beta_1) R_{t1} I ~ \delta_{i,1}.
\end{eqnarray}
It is much more convenient to introduce the matrix notation for the
algebraic manipulation. $|I \rangle  = ( I_1  I_2  I_3 )^{t}$, $|J
\rangle = ( J_1  J_2  J_3 )^{t}$, $|V \rangle = ( V_1 V_2 V_3
)^{t}$, and $|1 \rangle = ( 1  0  0 )^{t}$. Here the superscript $t$
represents the transpose of row vectors so that its effect is to
change them into column vectors. With these notations, Kirchhoff
rules can be written in compact forms as
\begin{eqnarray}
|V \rangle  &=& -(1 - \beta_1 \gamma_1) R_{t1} I ~ |1 \rangle
       + [ \hat{\beta} {\bf R} + \hat{\gamma} {\bf R}_t ] ~ |I \rangle  \nonumber\\
    &&  - \beta R {\bf A} |J \rangle , \\
|J \rangle  &=& -\frac{1}{2} |I \rangle  + \frac{1}{2} (\beta - \beta_1) I |1 \rangle , \\
0 &=& ({\bf R} + {\bf R}_t ) |I \rangle  - R {\bf A} |J \rangle   \nonumber\\
  && + (\beta_1 - \gamma_1) R_{t1} I ~ |1 \rangle .
\end{eqnarray}
Here ${\bf R}$ and ${\bf R}_t$ are diagonal matrices with diagonal elements $R_i$ and $R_{ti}$, respectively.
Similarly, $\hat{\beta}$ and $\hat{\gamma}$ are diagonal matrices with diagonal elements representing
the spin polarization of each FM electrode and the junction resistance, respectively.
${\bf A}$ is the matrix with its elements given by $A_{ij}$.
Formally, the unknown parameters can be written in a more compact matrix form as
\begin{eqnarray}
|I \rangle  &=& (\beta - \beta_1)I |1 \rangle   \nonumber\\
    && + [(\beta_1 - \beta) R_1 + (\gamma_1-\beta) R_{t1}] I {\bf G} |1 \rangle ,  \\
|J \rangle  &=& -\frac{1}{2} [(\beta_1 - \beta) R_1 + (\gamma_1-\beta) R_{t1}] I {\bf G} |1 \rangle ,  \\
|V \rangle  &=& - [ (\beta - \beta_1)^2 R_1 + (\beta^2 - 2\beta \gamma_1 +1) R_{t1}] I |1 \rangle   \nonumber\\
    && + [(\beta_1 - \beta) R_1 + (\gamma_1 - \beta) R_{t1}]  \nonumber\\
    && \times [ (\hat{\beta} - \beta) {\bf R} + (\hat{\gamma} - \beta) {\bf R}_t ] I ~ {\bf G} |1 \rangle .
\end{eqnarray}
Here the matrix ${\bf G}$ is defined by the expression
\begin{eqnarray}
\label{cond_matrix}
{\bf G} &=& \left[ {\bf R} + {\bf R}_t +\frac{1}{2} R {\bf A}  \right]^{-1}.
\end{eqnarray}
The matrix ${\bf G}$, with the dimension of conductance, is independent of magnetization configurations
(parallel or antiparallel to the spin-injecting electrode F1) of the FM electrodes.
The set of parameters, $I_i, J_i$ and $V_i$, contains all the information about the spin polarized
transport in nonlocal spin valves. In components, we have the spin currents
\begin{eqnarray}
\frac{I_i}{I}
  &=& (\beta-\beta_1) \delta_{i,1} \nonumber\\
  && + G_{i1} [ (\beta_1 - \beta) R_1 + (\gamma_1 - \beta) R_{t1}],  \\
\frac{J_i}{I}
  &=& -\frac{1}{2} G_{i1} [ (\beta_1 - \beta) R_1 + (\gamma_1 - \beta) R_{t1} ],
\end{eqnarray}
and the voltage drops in the FM electrodes
\begin{eqnarray}
\frac{V_i}{I}
  &=& - [ (\beta - \beta_1)^2 R_1 + (\beta^2 - 2\beta \gamma_1 +1) R_{t1}] \delta_{i,1} \nonumber\\
    && + [ (\beta_i - \beta) R_i + (\gamma_i - \beta) R_{ti} ] \nonumber\\
    && \times  G_{i1}
         [(\beta_1 - \beta) R_1 + (\gamma_1 - \beta) R_{t1}].
\end{eqnarray}
Note that the final results are written down in a very compact form, using new material parameters
as well as the conductance matrix.

\section{Physical meaning of spin resistance}
\label{s_resist}
 Spin resistance was defined in order to simplify the algebra.
In this section we are going to infuse some physical meaning into spin resistance.
Let us recast Eq.~(\ref{spin_R}) for $i\neq 1$ (nonlocal voltage probes) into a more illuminating form as
\begin{eqnarray}
\label{spin_R2}
\pm \frac{U_i}{1\pm\beta} - V_i
  &=& \pm \frac{I_i}{2} \left( \frac{2{\cal R}_{ti}}{1\pm\gamma_i} + \frac{2{\cal R}_i}{1\pm\beta_i} \right),
\end{eqnarray}
where $U_i$ acts as the effective electric potential of the base electrode at the junction with F$i$
and is defined by
\begin{eqnarray}
U_i &=& {\cal R} \sum_j J_j e^{-|L_i - L_j|/\lambda}.
\end{eqnarray}
For electrons with negative charge, $V_i$ is the electric potential
for both spin directions far into F$i$, and $\pm U_i/(1\pm \beta)$
is the electric potential for spin-up and spin-down electrons,
respectively, of the base electrode at the junction with F$i$. Refer
to Eqs.~(\ref{ecp_FM}) and (\ref{ecp_N}). We can deduce that the
current $I_i/2$ at the interface flows into (out of) F$i$ for
spin-up (spin-down) electrons.

 The left hand side (LHS) of Eq.~(\ref{spin_R2}) represents
 the electric potential difference for both spin directions
between the base electrode and F$i$ at the deep inside.
The right hand side (RHS) is the product of the current $I_i/2$ and
the effective spin-dependent resistance.
The sign in front represents correctly the flowing direction of the
spin-up and spin-down current, respectively.
The first term in the parenthesis is the spin-dependent tunnel
resistance as defined in Eq.~(\ref{R_ts}).
The second term is none other than the spin resistance,
which was introduced in the main text.
With this spin resistance, Eq.~(\ref{spin_R2}) is the effective
Ohm's law for the leakage spin-up
and spin-down currents.

 When $U_i$ is eliminated from Eq.~(\ref{spin_R2}), the relation between the nonlocal voltage drop $V_i$
and the leakage spin current $I_i$ or Eq.~(\ref{nonlocal_V2}) is obtained.
From Eq.~(\ref{spin_R2}), we can deduce the physical meaning of $V_i$.
No charge current flows in the nonlocal FM electrodes.
$V_i$ represents the shift of the electrochemical potential in F$i$
to satisfy the constraint of no charge current flow.
If we eliminate $V_i$ from Eq.~(\ref{spin_R2}), we find the following relation
\begin{eqnarray}
I_i (R_{ti} + R_i + \frac{R}{2} )
  &=& R \sum_{j\neq i} J_j e^{-|L_i - L_j|/\lambda}.
\end{eqnarray}
The material parameters are defined in Appendix \ref{dalgebra}. How do we interpret this relation?
This relation can be considered as the Ohm's law for the leakage spin current.
The LHS is the product of the spin current $I_i$ and the effective resistance.
From the standpoint of F$i$, the spin current $I_i$ flows from both sides of the base electrode ($R$)
through the junction ($R_{ti}$) and into F$i$ ($R_i$).
Hence the effective resistance is $R/2 + R_{ti} + R_i$ as in the above equation.
The RHS is the effective spin potential which combines the source term from F1 and
the sink terms from other nonlocal FM electrodes.

\section{Dependence on magnetization directions: three-dimensional case}
\label{3D-case}
One of key results of our paper is the independence
of the spin accumulation, the spin current, and the nonlocal voltage
on the magnetization directions of electrodes. This appendix is
aimed to provide an insight into the origin of this independence in
three-dimensional situations. We again consider the geometry in
Fig.~1. Similar notations will be used. The three-dimensional SDD
equation is given by Eq.~(\ref{SDDE_3d}) and
the associated charge and spin current densities ${\bf j}^{\rm c}$, ${\bf j}^{\rm s}$ are given by
\begin{eqnarray}
{\bf j}^{\rm c} & =& {1\over e}{\bf\nabla}(\sigma_+ \mu_+ + \sigma_- \mu_-), \\
{\bf j}^{\rm s} & =& {1\over e}{\bf\nabla}(\sigma_+ \mu_+ - \sigma_- \mu_-).
\end{eqnarray}
Similar relations hold for $\mu_{i\alpha}$, ${\bf j}^{\rm c}_i$, and ${\bf j}^{\rm s}_i$
in the F$i$ electrode.
The system is subject to the following boundary conditions.
From the condition of no leakage current to air or insulating substrate,
\begin{equation}
\hat{\bf n}\cdot \nabla \mu_\alpha=\hat{\bf n}_i\cdot \nabla \mu_{i\alpha}=0
\label{3D-no-leakage-current}
\end{equation}
should hold at the sample boundaries facing air or insulating substrate.
Here $\hat{\bf n}$ and $\hat{\bf n}_i$ denote normal vectors perpendicular to the boundaries.
From the constraint of the current continuity applied to the interface between the base
electrode and the electrode F$i$, one finds that the following relation should hold at the interface,
\begin{equation}
\hat{\bf n}\cdot \nabla \sigma_\alpha \mu_\alpha = \hat{\bf n} \cdot \nabla \sigma_{i\alpha} \mu_{i\alpha}.
\label{3D-current-continuity}
\end{equation}
The Ohm's law provides another boundary condition for the interface,
\begin{equation}
{1 \over e}\Delta \mu_{i\alpha}={\cal R}_{ti\alpha}\hat{\bf n}_i \cdot \nabla
{\sigma_{i\alpha}\mu_{i\alpha} \over e}.
\label{3D-Ohms-law}
\end{equation}
When combined with these boundary conditions, the SDD equation
completely fixes the spin-dependent electrochemical potentials. Here
we remark that the current density
$\nabla(\sigma_{i\alpha}\mu_{i\alpha}/e)$, instead of the current
$I_{i\alpha}$, appears in Eq.~(\ref{3D-Ohms-law}) and thus ${\cal
R}_{ti\alpha}$ in Eq.~(\ref{3D-Ohms-law}) amounts to the
spin-dependent junction resistance {\it per unit area}, instead of
the junction resistance. We also remark that the tunneling barrier
at a junction may not be uniform in realistic experimental
situations and such non-uniformity can be taken into account by
simply regarding ${\cal R}_{ti\alpha}$ as a position-dependent
quantity since Eq.~(\ref{3D-Ohms-law}) remains still valid even for
the nonuniform barrier as long as the tunnelling current remains
perpendicular to the interface.

In order to examine the dependence of the spin accumulation on the
magnetization directions, we reexpress the involved equations in
terms of the spin accumulation $\mu_+ - \mu_-$ and the charge
potential $(\sigma_+ \mu_+ + \sigma_- \mu_-)/\sigma$. The SDD
equation~(\ref{SDDE_3d}) is again decomposed into the following two
decoupled equations (spin mode and charge mode), \cite{selman}
\begin{eqnarray}
&& \nabla^2(\mu_+ - \mu_-)={1\over \lambda^2}(\mu_+ - \mu_-),
\label{3D-SDD-equation-spin-accumulation} \\
&& \nabla^2(\sigma_+ \mu_+ + \sigma_- \mu_-) = 0.
\label{3D-SDD-equation-charge-potential}
\end{eqnarray}
The boundary conditions for the spin accumulation can be derived
from Eq.~(\ref{3D-no-leakage-current}) and one obtains
\begin{equation}
\hat{\bf n}\cdot \nabla (\mu_+ - \mu_-) = \hat{\bf n}_i \cdot \nabla(\mu_{i+} - \mu_{i-})=0,
\label{3D-no-leakage-current-spin-accumulation}
\end{equation}
from Eq.~(\ref{3D-current-continuity}), one obtains
\begin{eqnarray}
&& \hat{\bf n}\cdot \nabla(\mu_{i+} - \mu_{i-})
\label{3D-current-continuity-spin-accumulation}
\\
&&={\sigma \over \sigma_i}{1-\beta^2 \over 2}\left(
{1 \over 1+\beta_i} + {1 \over 1-\beta_i} \right)
\hat{\bf n}\cdot \nabla(\mu_+ - \mu_-)
\nonumber \\
&& +{e \over \sigma_i} \left(
{1+\beta \over 1+\beta_i} - {1-\beta \over 1-\beta_i} \right)
\hat{\bf n}\cdot {\bf j}^{\rm c}_i,
\nonumber
\end{eqnarray}
and from Eq.~(\ref{3D-Ohms-law}), one obtains
\begin{eqnarray}
&& {1\over e}\Delta(\mu_{i+}-\mu_{i-})
\label{3D-Ohms-law-spin-accumulation}
\\
&&
={\sigma_i {\cal R}_{ti}(1-\beta_i^2) \over 4e} \hat{\bf n}\cdot \nabla (\mu_{i+}-\mu_{i-})
+ {{\cal R}_{ti}(\beta_i+\gamma_i) \over 2} \hat{\bf n}\cdot {\bf j}^{\rm c}_i.
\nonumber
\end{eqnarray}

Now we are ready to discuss the magnetization direction dependence of the spin accumulation,
which is completely fixed from its SDD equation (\ref{3D-SDD-equation-spin-accumulation})
and boundary conditions~(\ref{3D-no-leakage-current-spin-accumulation}),
(\ref{3D-current-continuity-spin-accumulation}),
and (\ref{3D-Ohms-law-spin-accumulation}).
Note that in these equations, all terms that depend on the magnetization directions
are multiplied by the charge current density. Thus the spin accumulation should be
independent of the magnetization directions of electrodes in which the charge current density vanishes.

Next we discuss the magnetization direction dependence of the spin current.
The spin current can be obtained from the spin accumulation as follows,
\begin{eqnarray}
{\bf j}^{\rm s} & = & {\sigma(1-\beta^2) \over 2e} \nabla(\mu_+ - \mu_-)
+\beta {\bf j}^{\rm c},
\\
{\bf j}^{\rm s}_i & = & {\sigma_i(1-\beta^2_i) \over 2e} \nabla(\mu_{i+} - \mu_{i-})
+\beta {\bf j}^{\rm c}_i.
\end{eqnarray}
Then from the properties of the spin accumulation, it is evident that
the spin current density should be independent of the magnetization directions
of electrodes in which the charge current density vanishes.

Finally we discuss the charge potential, which is subject to the SDD
equation (\ref{3D-SDD-equation-charge-potential}). The boundary
conditions for the charge potential can be derived from
Eqs.~(\ref{3D-no-leakage-current}), and one obtains
\begin{equation}
\hat{\bf n}\cdot \nabla(\sigma_+ \mu_+ + \sigma_- \mu_-)
=\hat{\bf n}_i\cdot \nabla(\sigma_{i+} \mu_{i+} + \sigma_{i-} \mu_{i-})=0,
\label{3D-no-leakage-current-charge-potential}
\end{equation}
from Eq.~(\ref{3D-current-continuity}), one obtains
\begin{equation}
\hat{\bf n}\cdot \nabla (\sigma_+ \mu_+ + \sigma_- \mu_-)
=\hat{\bf n}\cdot \nabla (\sigma_{i+} \mu_{i+} + \sigma_{i-} \mu_{i-}),
\label{3D-current-continuity-charge-potential}
\end{equation}
and from Eq.~(\ref{3D-Ohms-law}), one obtains
\begin{eqnarray}
&& {1\over e}{\sigma_{i+}\mu_{i+} + \sigma_{i-}\mu_{i-} \over \sigma_i}
-{1\over e}{\sigma_{+}\mu_{+} + \sigma_{-}\mu_{-} \over \sigma}
\label{3D-Ohms-law-charge-potential}
\\
&& ={\beta_i - \beta \over 2e}(\mu_+ - \mu_-)
\nonumber \\
&& + {\sigma_i {\cal R}_{ti}(1-\beta_i^2) \over 16e}\left[
(1+\beta_i)(1+\gamma_i)-(1-\beta_i)(1-\gamma_i)\right]
\nonumber \\
&& \,\,\,\,\,\, \times
\hat{\bf n}\cdot \nabla (\mu_{i+}-\mu_{i-})
\nonumber \\
&& + { {\cal R}_{ti} \over 8}\left[ (1+\beta_i)^2(1+\gamma_i)+(1-\beta_i)^2(1-\gamma_i)\right]
\hat{\bf n}\cdot {\bf j}^{\rm c}_i.
\nonumber
\end{eqnarray}
Note that the SDD equation~(\ref{3D-SDD-equation-charge-potential}) and the boundary
conditions~(\ref{3D-no-leakage-current-charge-potential}), (\ref{3D-current-continuity-charge-potential})
are not dependent on the magnetization directions of any electrodes.
Thus the magnetization direction dependence can arise only from the
boundary condition~(\ref{3D-Ohms-law-charge-potential}).
From Eq.~(\ref{3D-Ohms-law-charge-potential}) and from the properties
of the spin accumulation, one then finds that the charge potential
at the electrode F$i$ is independent of the magnetization direction of
other non-current-carrying electrodes.
This in turn implies that the nonlocal voltage measured between the
electrode F$i$ and the base electrode ($x=+\infty$) should be independent
of the magnetization directions of other non-current-carrying
electrodes F$j$ $(j\neq i)$.


\begin{thebibliography}{100}

\bibitem{spinreview} I. Zutic, J. Fabian, S. Das Sarma, Rev. Mod. Phys. {\bf 76}, 323 (2004).

\bibitem{storque1} L. Berger, Phys. Rev. B {\bf 54}, 9353 (1996).
\bibitem{storque2} J. Slonczewski, J. Magn. Magn. Mater. {\bf 159}, L1 (1996).
\bibitem{storque3} For a recent review, see A. Bratass, G. E. W. Bauer, and P. J. Kelly,
  Phys. Rep. {\bf 427}, 157 (2006);
  Y. Tserkovnyak, A. Brataas, G. E. W. Bauer, and B. I. Halperin,
  Rev. Mod. Phys. {\bf 77}, 1375 (2005).

\bibitem{gmr1} M. N. Baibich, J. M. Broto, A. Fert, F. Nguyen Van Dau, F. Petroff, P. Etienne, G. Creuzet,
  A. Friederich, and J. Chazelas,
  Phys. Rev. Lett. {\bf 61}, 2472 (1988).
\bibitem{gmr2} G. Binasch, P. Gr\"{u}nberg, F. Saurenbach, and W. Zinn,
  Phys. Rev. B {\bf 39}, 4828 (1989).

\bibitem{mtj} For a recent review, see E. Y. Tsymbal, O. N. Mryasov, and P. R. LeClair,
 J. Phys.: Condens. Matter {\bf 15}, 109 (2003).

\bibitem{nanopillar} J. A. Katine, F. J. Albert, R. A. Buhrman, E. B. Myers, and D. C. Ralph,
  Phys. Rev. Lett. {\bf 84}, 3149 (2000).


\bibitem{johnson0} M. Johnson and R. H. Silsbee,
  Phys. Rev. Lett. {\bf 55}, 1790 (1985);
  Phys. Rev. B {\bf 35}, 4959 (1987);
  \textit{ibid} {\bf 37}, 5312 (1988).


\bibitem{jedema}
 F. J. Jedema, H. B. Heersche, A. T. Filip, J. J. A. Baselmans, and B. J. van Wees,
  Nature (London) {\bf 416}, 713 (2002);
 F. J. Jedema, M. S. Nijboer, A. T. Filip, and B. J. van Wees,
  Phys. Rev. B {\bf 67}, 085319 (2003).

\bibitem{ji} Y. Ji, A. Hoffmann, J. S. Jiang, J. E. Pearson, and S.
D. Bader, J. Phys. D: {\bf 40}, 1280 (2007).

\bibitem{tinkham_SHE} S. O. Valenzuela and M. Tinkham, Nature {\bf 442}, 176 (2006);
  J. Appl. Phys. {\bf 101}, 09B103 (2007).
\bibitem{otani_SHE} T. Kimura, Y. Otani, T. Sato, S. Takahashi, and S. Maekawa,
  Phys. Rev. Lett. {\bf 98}, 156601 (2007).


\bibitem{otani} T. Kimura, J. Hamrle, Y. Otani, K. Tsukagoshi, and Y. Aoyagi,
  Appl. Phys. Lett. {\bf 85}, 3795 (2004).

\bibitem{otani2} T. Kimura, J. Hamrle, and Y. Otani, Phys. Rev. B {\bf 72}, 014461 (2005).

\bibitem{johnson} R. Godfrey and M. Johnson, Phys. Rev. Lett. {\bf 96}, 136601 (2006).

\bibitem{wees} M. V. Costache, M. Zaffalon, and B. J. van Wees,
  Phys. Rev. B {\bf 74}, 012412 (2006).




\bibitem{son} P. C. van Son, H. van Kempen, and P. Wyder, Phys. Rev. Lett. {\bf 58}, 2271 (1987).

\bibitem{brataas} A. Brataas, Y. V. Nazarov, and G. E. W. Bauer,
Eur. Phys. J. B {\bf 22}, 99 (2001).

\bibitem{kovalev} A. A. Kovalev, A. Brataas, and G. E. W. Bauer,
Phys. Rev. B {\bf 66}, 224424 (2002).

\bibitem{kovalev2} A. A. Kovalev, G. E. W. Bauer, and A. Brataas,
Phys. Rev. B {\bf 73}, 054407 (2006).

\bibitem{valet_fert} T. Valet and A. Fert, Phys. Rev. B {\bf 48}, 7099 (1993).

\bibitem{penn_stiles} D. R. Penn and M. D. Stiles, Phys. Rev. B {\bf 72}, 212410 (2005).

\bibitem{selman} S. Hershfield and H. L. Zhao, Phys. Rev. B {\bf 56}, 3296 (1997).

\bibitem{takahashi} S. Takahashi and S. Maekawa, Phys. Rev. B {\bf 67}, 052409 (2003).

\bibitem{bclee} B. C. Lee, T.-S. Kim, K. Rhie and J. Hong, Appl. Phys. Lett. {\bf 91}, 022504 (2007).

\bibitem{hamrle} J. Hamrle, T. Kimura,
 Y. Otani, K. Tsukagoshi, and Y. Aoyagi, Phys. Rev. B {\bf 71}, 094402 (2005).


\bibitem{Bass} J. Bass and W. P. Pratt Jr., J. Magn. Magn. Mater. {\bf 200}, 274 (1999).
\bibitem{Fert} A. Fert and L. Piraux, J. Magn. Magn. Mater. {\bf 200}, 338 (1999).


\bibitem{mr_cnt1} S. Sahoo, T. Kontos, J. Furer, C. Hoffmann, M. Gr\"{a}ber, A. Cottet, 
and C. Sch\"{o}nenberger, Nature Phys. {\bf 1}, 99 (2005).

\bibitem{mr_cnt2} H. T. Man, I. J. W. Wever, and A. F. Morpurgo, 
  Phys. Rev. B {\bf 73}, 241401(R) (2006). 

\bibitem{BNB_circuit} Our results for MR, in fact, are in agreement with the collinear case of A. Brataas, Yu. V. Nazarov, and G. E. W. Bauer,
  Phys. Rev. Lett. {\bf 84}, 2481 (2000).  
\bibitem{tskim} T.-S. Kim, {\it unpublished work} (2008).


\end{thebibliography}
\end{document}